\documentclass[letterpaper, 10 pt, conference]{ieeeconf}  

\usepackage{graphicx}
\usepackage{multirow}
\usepackage{colortbl}
\usepackage[table,xcdraw]{xcolor}
\usepackage{colortbl}
\usepackage{url}
\usepackage{pifont}
\usepackage{amsmath}

\IEEEoverridecommandlockouts                              

\title{\LARGE \bf
Toward Zero-Shot Learning for Visual Dehazing of Urological Surgical Robots
}

\author{Renkai Wu$^{1,*}$, Xianjin Wang$^{3,*}$, Pengchen Liang$^{1}$, Zhenyu Zhang$^{4}$, Qing Chang\textsuperscript{1,\ding{41}} and Hao Tang\textsuperscript{2,\ding{41}}
\thanks{\ding{41} Corresponding authors.}
\thanks{$^{1}$Department of Surgery, Shanghai Key Laboratory of Gastric Neoplasms, Shanghai Institute of Digestive Surgery, Ruijin Hospital, Shanghai Jiao Tong University School of Medicine, Shanghai, China.
        {\tt\small wurk@shu.edu.cn, liangpengchen@shu.edu.cn, robie0510@hotmail.com}}%
\thanks{$^{2}$National Key Laboratory for Multimedia Information Processing, 
School of Computer Science, Peking University, Beijing, China.
    {\tt\small haotang@pku.edu.cn}}%
\thanks{$^{3}$Department of Urology, Ruijin Hospital, Shanghai Jiaotong University School of Medicine, Shanghai, China.
        {\tt\small xianjin09@163.com}}
\thanks{$^{4}$ School of Intelligent Science and Technology, Nanjing University, China.
        {\tt\small zhenyuzhang@nju.edu.cn}}
\thanks{$^{*}$These authors contributed equally: Renkai Wu and Xianjin Wang.}
}
\begin{document}

\maketitle
\thispagestyle{empty}
\pagestyle{empty}

\begin{abstract}

Robot-assisted surgery has profoundly influenced current forms of minimally invasive surgery. However, in transurethral suburethral urological surgical robots, they need to work in a liquid environment. This causes vaporization of the liquid when shearing and heating is performed, resulting in bubble atomization that affects the visual perception of the robot. This can lead to the need for uninterrupted pauses in the surgical procedure, which makes the surgery take longer. To address the atomization characteristics of liquids under urological surgical robotic vision, we propose an unsupervised zero-shot dehaze method (RSF-Dehaze) for urological surgical robotic vision. Specifically, the proposed Region Similarity Filling Module (RSFM) of RSF-Dehaze significantly improves the recovery of blurred region tissues. In addition, we organize and propose a dehaze dataset for robotic vision in urological surgery (USRobot-Dehaze dataset). In particular, this dataset contains the three most common urological surgical robot operation scenarios. To the best of our knowledge, we are the first to organize and propose a publicly available dehaze dataset for urological surgical robot vision. The proposed RSF-Dehaze proves the effectiveness of our method in three urological surgical robot operation scenarios with extensive comparative experiments with 20 most classical and advanced dehazing and image recovery algorithms. The proposed source code and dataset are available at~\url{https://github.com/wurenkai/RSF-Dehaze} .

\end{abstract}

\section{Introduction}

Recently, robot-assisted surgery has profoundly changed the current approach to minimally invasive surgery (MIS) by expanding MIS indications with the help of high robotic dexterity and high-precision motion accuracy \cite{luciani2017robotic,hyams2008robotic,cohen2015robotic}. Specifically, in urology, in May 2018 the US Food and Drug Administration (FDA) approved the da Vinci single port (SP) system \cite{gosrisirikul2018new} for use in urology. The system is the first surgical system with a dedicated single point of view to hit the market. Clinicians have reported their effectiveness with the da Vinci SP in surgical scenarios such as radical prostatectomy, radical, and partial nephrectomy \cite{agarwal2020initial,moschovas2020technical,kaouk2019step,zhang2020single}.

\begin{figure}[!t]
\centering
\includegraphics[width=\linewidth]{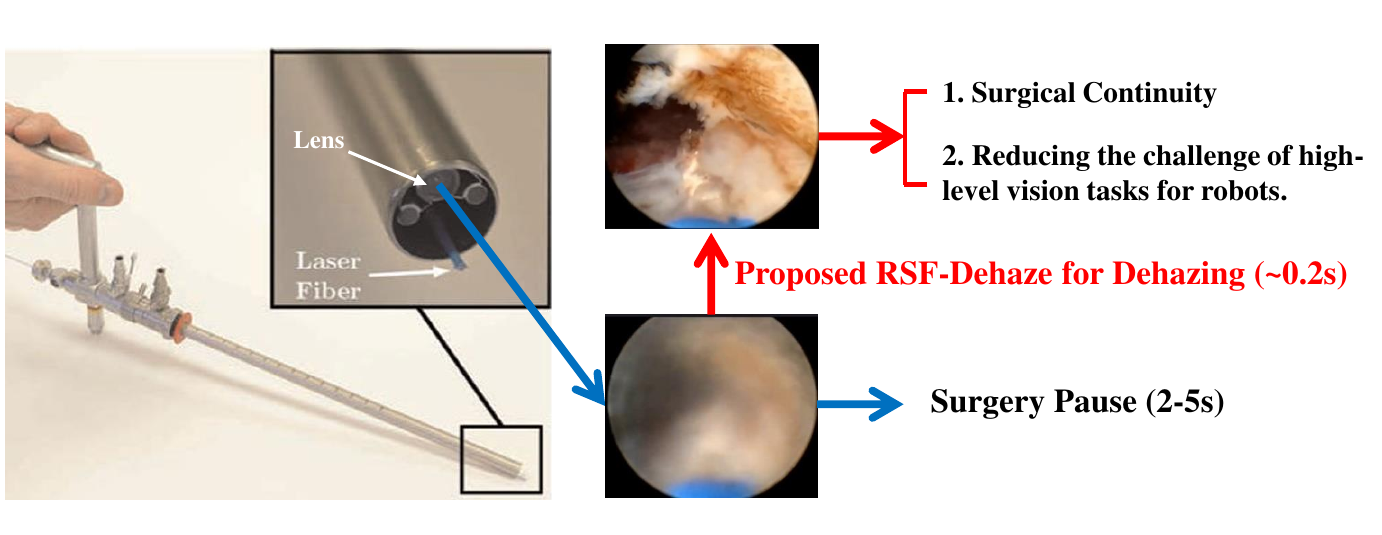}
\caption{Images of the transurethral urology surgical robot (robot image source Hendrick et al. \cite{hendrick2015hand}). In HoLEP surgery, the robot is introduced through the urethra and a laser fiber is used to remove tissue.}
\label{fig00}
\end{figure}

However, some transurethral urologic procedures \cite{mitchell2016toward,hendrick2015hand,amanov2020transurethral} are performed through the most demanding orifice (urethra) into the body. This has allowed SP systems to become more miniaturized (see Fig. \ref{fig00}) which makes the vision obtained by the clinician narrower. In particular, due to the specific nature of transurethral urological surgery, cutting tissue through a laser in a liquid environment such as the urethra will create a large number of air bubbles \cite{hendrick2015hand,amanov2020transurethral}. This causes the field of view to become blurred and the clinician performing a robotic-assisted procedure with such a restricted field of view needs to stop and wait for the bubble-induced atomization blur to dissipate. This waiting process usually takes a few seconds, which increases the procedure time and stress for both the clinician and the patient. Therefore, effective removal of the bubble atomization blur of urological surgical robotic vision in a liquid environment has become a critical requirement.

With the rapid development of computer technology, fog removal and image restoration through computer vision techniques have become a common method \cite{wang2022brief,guo2023haze,goyal2023recent}. These methods include prior-based, supervised, or unsupervised fog removal methods. The early prior-based methods are FVR \cite{tarel2009fast}, DCP \cite{he2010single} and BCCR \cite{meng2013efficient}. Supervised methods include AOD-Net \cite{li2017aod}, GCANet \cite{chen2019gated}, 4KDehazing \cite{zheng2021ultra}, Dehazeformer \cite{song2023vision}, Histoformer \cite{sun2024restoring}, and KA-Net \cite{feng2024advancing}. Unsupervised methods include DIP \cite{ulyanov2018deep}, DD \cite{heckel2018deep}, DDIP \cite{gandelsman2019double}, ZID \cite{li2020zero}, YOLY \cite{li2021you}, USID-Net \cite{li2022usid}, D4 \cite{yang2022self}, ZS-NSN \cite{mansour2023zero}, UCL-Dehaze \cite{wang2024ucl} and DIPDKP \cite{yang2024dynamic}. However, these methods are designed based on the fog of natural scenes, which often shows degraded performance used in the liquid environment of urological surgical robotic vision. This is due to the fact that the robotic vision of urologic surgery differs significantly from images of the natural environment, including dim light, a narrow field of view, and execution in a liquid environment. All of these factors affect light scattering in vision.

\begin{figure}[!t]
\centering
\includegraphics[width=\linewidth]{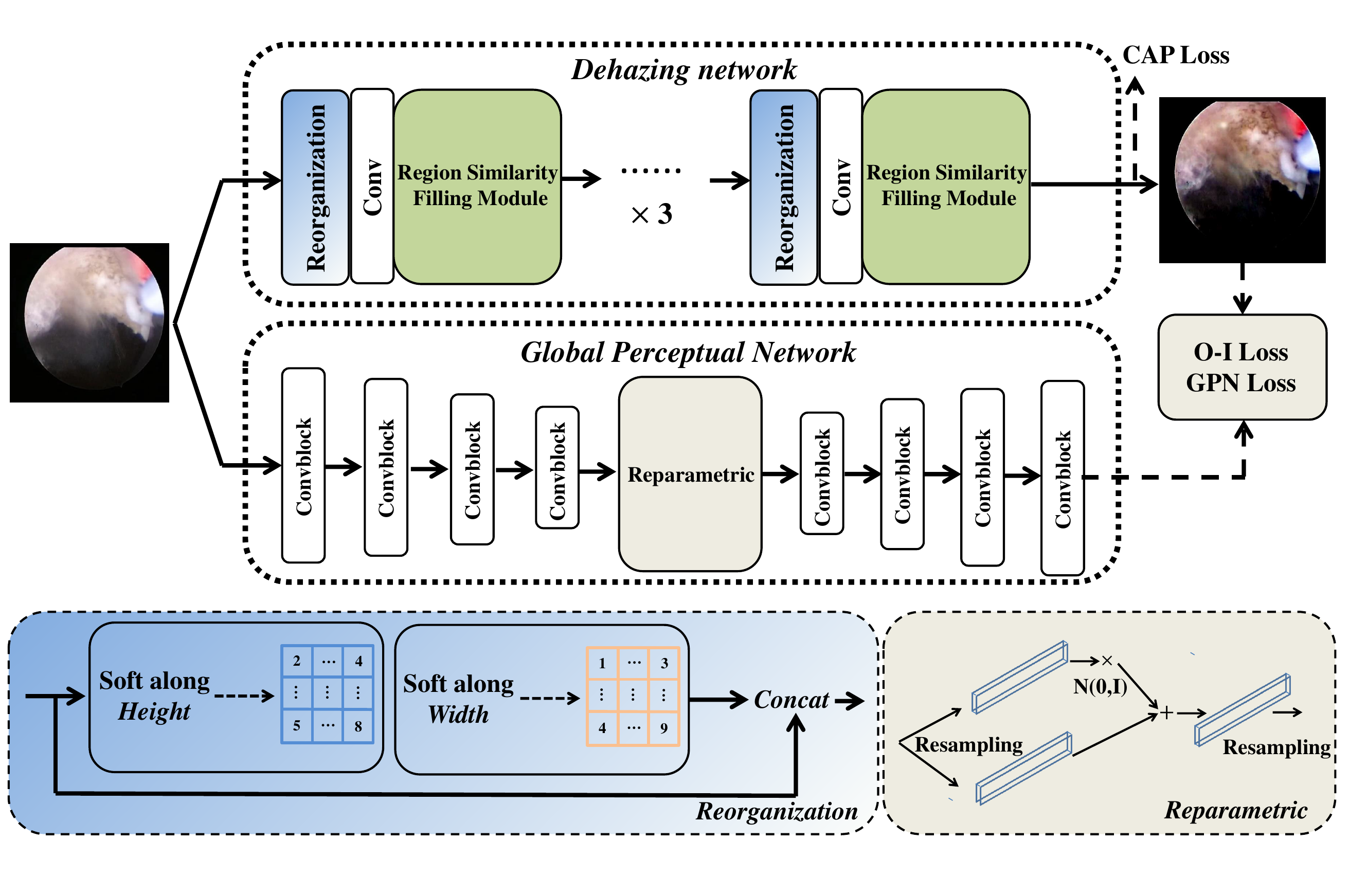}
\caption{Our proposed unsupervised zero-shot dehazing method (RSF-Dehaze) for dehaze processing of robotic vision for urological surgery.}
\label{fig02}
\end{figure}

Therefore, in order to address the problem of fog removal in robotic vision for urological surgery, we propose an unsupervised zero-shot dehaze method (RSF-Dehaze) to address this challenge (see Fig. \ref{fig02}). 
Supervised methods require pair-wise clean and fuzzy data for training; however, this is often difficult to obtain. And the use of synthesized fogged and real images inevitably reduces the model's ability to cope with real scenes \cite{wang2024ucl}. Some unsupervised contrastive learning approaches employ unpaired data for training; however, their effectiveness is highly dependent on the amount of unpaired data \cite{li2021you}. Therefore, we propose an unsupervised zero-shot dehazing method that learns and performs dehazing from only a single input image. Specifically, we convert the image to a YCbCr space based on the visual characteristics of a urological surgical robot. Then, we design a region similarity filling module to recover the regions blurred by the fog produced by bubbles based on the characteristics of bubble-generated fog. All processes are learned and processed by input single image only. 

In addition, to deal with the problem that there is no publicly available robotic vision dehazing dataset for urological surgery, we organized and proposed dehazing datasets for different urological surgical scenarios in three real-life scenarios. Specifically, these include Green Laser Prostate Vaporization, Prostate Plasma Enucleation, and Holmium Laser Prostatectomy. We name the proposed dataset as USRobot-Dehaze dataset and make it publicly available to accelerate the development in the field of robotic vision for urological surgery.

\begin{itemize}
\item We propose an unsupervised zero-shot dehazing method (RSF-Dehaze) for dehazing processing of urological surgery robotic vision. Specifically, we propose a region similarity filling module based on the bubble-generated fogging of urological surgery robotic vision. RSF-Dehaze is a novel unsupervised zero-shot dehazing method that only requires fog removal from a single urological surgery robotic vision image.

\item We organized and proposed dehazing datasets for three real-life urological surgical scenarios and name them USRobot-Dehaze dataset. In particular, the proposed USRobot-Dehaze dataset solves this problem that there is no publicly available visual dehazing dataset for robots in urological surgery.

\item Through comparative experiments under three different urological surgical scenarios in real-life scenarios. The proposed RSF-Dehaze is fully compared with 20 representative state-of-the-art dehazing and image recovery methods to confirm its effectiveness.

\item The source code and propoosed dataset are publicly available at our GitHub.

\end{itemize}

\begin{figure*}[!t]
\centering
\includegraphics[width=\linewidth]{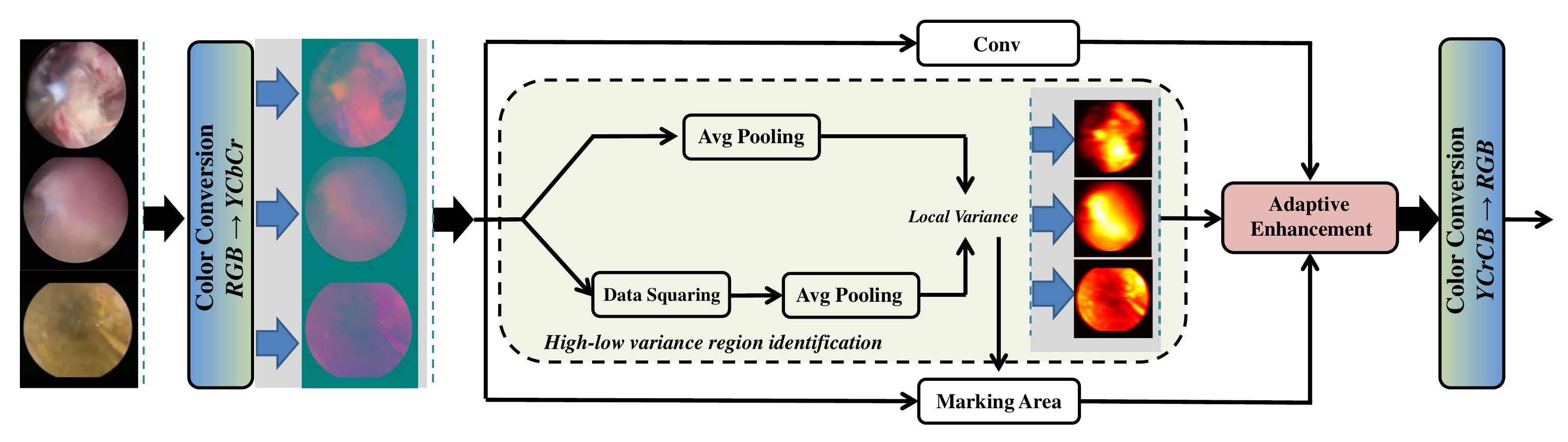}
\caption{Component structure of the proposed Regional Similarity Filling Module (RSFM).}
\label{fig03}
\end{figure*}

\section{Methodology}

\subsection{The Proposed Method}
\subsubsection{Architecture overview}
Our proposed unsupervised zero-shot dehaze model (RSF-Dehaze) for robotic vision in urological surgery is shown in Fig. \ref{fig02}. Specifically, RSF-Dehaze consists of two networks for composition, including the dehazing network and the global perception network. The dehazing network serves as the cornerstone of our approach, comprising five identical composite units. Each unit consists of three modules: a reorganization module, a convolution module, and a region similarity filling module. The global perceptual network acts as sensing the global luminosity in the image, independent of the image content \cite{li2021you}. It mainly consists of an encoder, decoder, and intermediate blocks.

\subsubsection{Dehazing network}
Dehazing network is the key network of the proposed RSF-Dehaze. It is mainly composed of five identical combinations, including a reorganization module, a convolution, and a region similarity filling module. Specifically, we address the formation of fogging blur due to air bubbles in the vision of a urological surgical robot, thus proposing a Region Similarity Filling Module (RSFM) for dehazing. More specifically, the blurring of urological surgery robotic vision, as shown in Fig. \ref{fig00}, it is more due to air bubbles that result in localized regions of relative blurring in the image. However, urological surgical images are distinguished from images of natural scenes in that the field of view is very restricted, and often there is clear and identical tissue recovered in a blurred image. Therefore, we take advantage of this one feature and propose RSFM for dehazing blurring in robotic vision for urological surgery.

Specifically, the composition structure of the region similarity filling module (RSFM) is shown in Fig. \ref{fig03}. The blurred image input to the RSFM first undergoes a color space conversion to convert the initial RGB space to YCbCr space. This is because we found that in the liquid environment under robotic vision in urological surgery, the fogging effect in the liquid can be better reduced by being in the YCbCr space to improve the clarity to the real tissue. 

The image in YCbCr space after conversion goes through three separate paths. The second intermediate path is the critical part, specifically, the input images are subjected to local variance (frequency domain) computation, respectively. We perform this operation through an average pooling operation and a data squaring operation. As shown in Fig.~\ref{fig03}, we provide a visualization of the variance (frequency domain) heatmap. From the heat map, we can learn that a clear organization structure is a high-variance (frequency domain) information, while low-variance (frequency domain) information is fuzzy organization information. However, in robotic vision for urologic surgery, often within a restricted field of view, the tissue structure is usually similar. We take advantage of this feature to complement the information from the clear organizational structure of the high-variance (frequency domain) inside the fuzzy organization of the low-variance (frequency domain). This operation is performed in the adaptive enhancement operation after combining three paths. In the third path, we mark the variance map obtained through the computation in the original image. In addition, in the first path, we extract and further enhance the high-frequency information by using a high-pass filter. Then, the enhanced high frequency information, the marked image and the variance information are subjected to adaptive enhancement operations. Finally, the YCbCr space is converted again to the RGB space by color space conversion. The region similarity filling module (RSFM) operation can be expressed by the following equations:
\begin{equation}
\begin{aligned}
Y&=rgb\_ycbcr(X), \\
F&=AP[Sq(Y)]-[A p(Y)]^{2}, \\
Y\_L,Y\_H &= M(F,Y,T), \\
NY\_L &= Y\_L + Adp[Conv(Y),F],\\
Out &= ycbcr\_rgb(NY\_L,Y\_H),
\end{aligned}
\end{equation}
\noindent
where $rgb\_ycbcr$ and $ycbcr\_rgb$ are color space conversion operations, $AP$ is average pooling operation, and $Sq$ is squaring operation, respectively. The $M(x, y, z)$ is the region marking operation, where $x$ is the variance map, $y$ the original map, and $z$ is the threshold (default is 0.2 times the maximum variance). The $Y\_L$ and $Y\_H$ are the labeled low-variance and high-variance (frequency-domain) regions. The $Adp$ is the adaptive enhancement operation, which combines the high-frequency information after high-pass filtering ($Conv(Y)$) and the high-frequency information after high-pass filtering ($F$) and uses different intensity enhancement effects for different blurring regions.

In addition, to better stabilize the computation of variance (frequency domain) in the Regional Similarity Filling Module (RSFM), we employed pixel reorganization on the input RSFM image. As shown in Fig. \ref{fig02}, we do this by rearranging half of the channels first with the pixel values in the vertical direction and then with the pixel values in the horizontal direction. Finally, the rearranged image is concatenated with the original image. By doing this, the variance (frequency domain) distribution of the image can be significantly smoothed, and the subsequent error in calculating the local variance (frequency domain) can be reduced.

\subsubsection{Global perceptual network}
Inspired by Li et al. \cite{li2020zero}, we use a global perceptual network as shown in Fig. \ref{fig02} to perceive global luminosity. The global perception network consists mainly of encoders, decoders, and intermediate blocks. Among them, each encoder and decoder consists mainly of a convolutional layer, an activation function, and a pooling layer or upsampling. In particular, the learning of global luminosity by the global perceptual network is defined as a variational inference problem. In the intermediate block, the output of the encoder is transformed to a hidden Gaussian distribution to obtain a hidden Gaussian model. This includes obtaining the mean and variance of the hidden Gaussian model, and the reconstruction of the potential code is obtained by resampling. Finally, it is then fed into the decoder for global luminosity reconstruction.

\subsubsection{Loss function}
The loss function is mainly made up of the dehazing network and the global perceptual network. In Global Perception Network ($GPN Loss$) we keep it consistent with Li et al. \cite{li2021you}. In the dehazing network, we perform the CAP loss by taking the dehazing input. This is used to calculate the loss values for luminance and saturation. We learn this by using the following equations:
\begin{equation}
\mathcal{L}_{CAP} = MSE[B(x),S(x)],
\end{equation}
\noindent
where $B(x)$ represents the brightness of the output, $S(x)$ represents the saturation of the output, and $MSE$ represents the mean square error.

In addition, since the RSF-Dehaze model is an unsupervised zero-shot dehazing network, we also minimize the loss of the dehazing network outputs and inputs:
\begin{equation}
\mathcal{L}_{O-I} = MSE[D(x),x],
\end{equation}
\noindent
where $D(x)$ represents the clean image that has been processed by the dehazing network and $x$ is represents the input image.

\begin{figure}[!t]
\centering
\includegraphics[width=\linewidth]{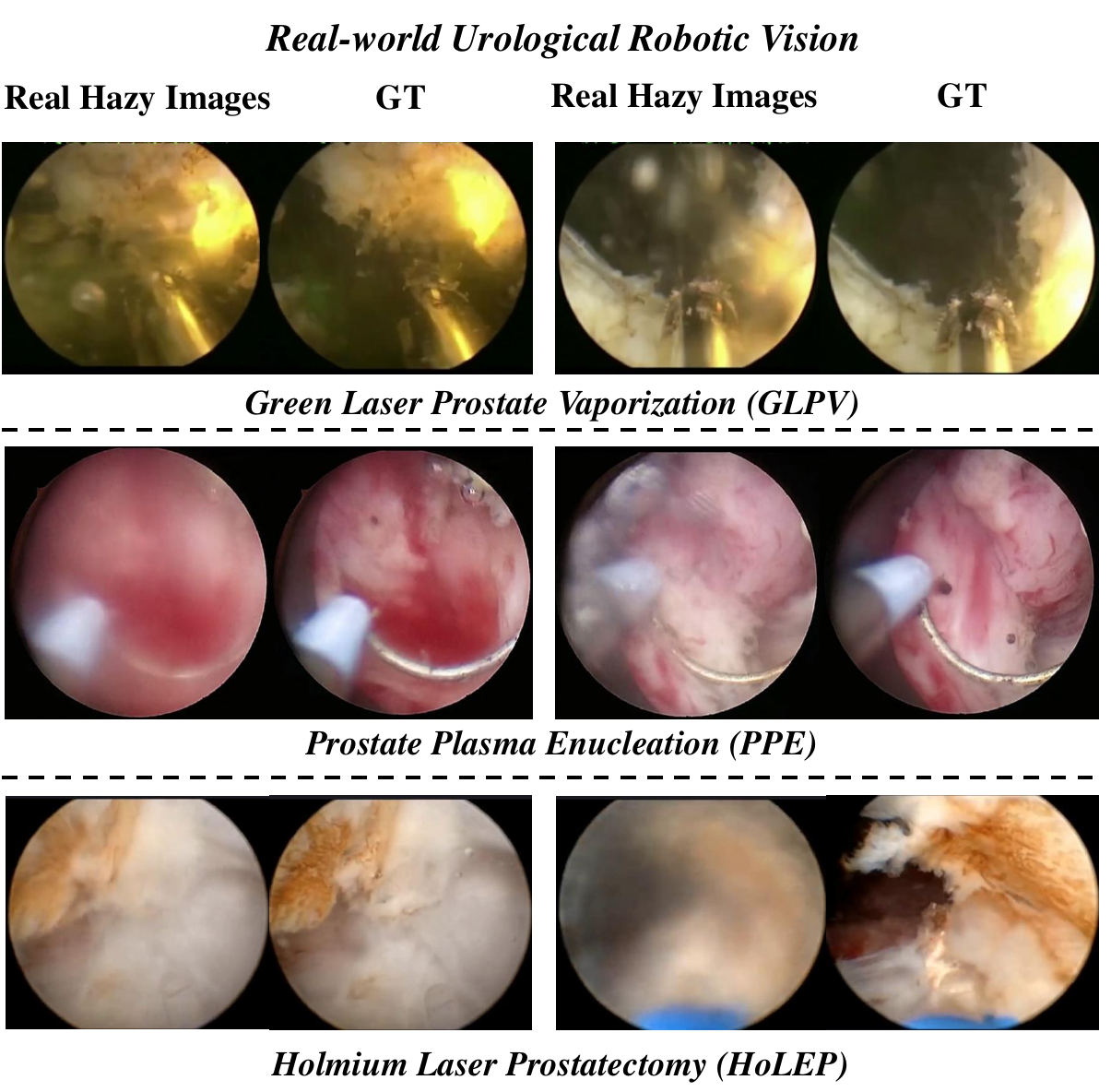}
\caption{The proposed dehaze dataset (USRobot-Dehaze) for three urological surgical robot operation scenarios.}
\label{fig01}
\end{figure}

\subsection{The Proposed Dataset}

In order to fill this gap of the current lack of dehazing datasets for urological surgical robot vision, we propose a real-scene based fog removal dataset for urological surgical robot vision and name it USRobot-Dehaze dataset. To the best of our knowledge, the proposed USRobot-Dehaze dataset is the first collated and publicly available visual dehazing dataset for urological surgical robots. Specifically, we first collected and organized currently publicly available videos of urology surgical robot surgical operations from the Internet. In particular, these include the three most common urological surgical scenarios and protocols, namely Green Laser Prostate Vaporization (GLPV), Prostate Plasma Enucleation (PPE), and Holmium Laser Prostateectomy (HoLEP). We extract key frames from the collected and organized surgical operation videos and collect a pair of blurred images and clean images in the same identical scene. More specifically, we extracted a total of 194 keyframes that were eventually collated to form 97 pairs of images with clean and blurred images from real scenes. We standardized the image size to 1080$\times$1080. We reclassified the USRobot-Dehaze dataset according to three surgical protocols: GLPV, PPE, and HoLEP, as illustrated in Fig. \ref{fig01}. The dataset comprises real-scene images from robotic urological surgeries, with 32 pairs for GLPV, 49 pairs for PPE, and 16 pairs for HoLEP. Each pair consists of a clean image and its corresponding blurred counterpart. 

\begin{table*}[!t]
\centering
\caption{Performance comparison with 20 most classical and state-of-the-art models on the USRobot-Dehaze dataset. The red color font represents the best performance and the orange color font represents the second best performance.}
\renewcommand{\arraystretch}{1.10}
\resizebox{\linewidth}{!}{
\begin{tabular}{lccccccccc}
\hline
                                   &                                        &                                 &                                           & \multicolumn{2}{c}{\textbf{GLPV}}                            & \multicolumn{2}{c}{\textbf{PPE}}                             & \multicolumn{2}{c}{\textbf{HoLEP}}                           \\ \cline{5-10} 
\multirow{-2}{*}{\textbf{Method}} & \multirow{-2}{*}{\textbf{Publication}} & \multirow{-2}{*}{\textbf{Type}} & \multirow{-2}{*}{\textbf{Inference Time}} & \textbf{PSNR}                & \textbf{SSIM}                 & \textbf{PSNR}                & \textbf{SSIM}                 & \textbf{PSNR}                & \textbf{SSIM}                 \\ \hline
FVR \cite{tarel2009fast}                                & ICCV'09                                & Prior                           & 0.17s                                     & 15.00                        & 0.5826                        & 13.35                        & 0.4896                        & 10.99                        & 0.6069                        \\
DCP \cite{he2010single}                               & TPAMI'10                               & Prior                           & 0.22s                                     & 16.06                        & 0.7531                        & 12.45                        & 0.6191                        & 11.19                        & 0.6859                        \\
BCCR \cite{meng2013efficient}                              & ICCV'13                                & Prior                           & 0.26s                                     & 13.00                        & 0.6076                        & 13.26                        & 0.5534                        & 12.29                        & 0.6662                        \\ \hline
AOD-Net \cite{li2017aod}                           & ICCV'17                                & Supervised                      & 0.18s                                     & 16.26                        & 0.7174                        & 14.39                        & 0.5744                        & 14.31                        & 0.7760                        \\
GCANet \cite{chen2019gated}                            & WACV'19                                & Supervised                      & 0.51s                                     & 19.28                        & 0.7962                        & 16.13                        & 0.6851                        & 19.58                        & 0.8483                        \\
4KDehazing \cite{zheng2021ultra}                        & CVPR'21                                & Supervised                      & 0.11s                                     & 15.38                        & 0.4734                        & 15.53                        & 0.4923                        & 17.07                        & 0.6513                        \\
Dehazeformer-t \cite{song2023vision}                    & TIP'23                                 & Supervised                      & 0.23s                                     & 18.78                        & 0.8440                        & 15.83                        & 0.6603                        & 20.49                        & 0.8997                        \\
Dehazeformer-m \cite{song2023vision}                    & TIP'23                                 & Supervised                      & 0.75s                                     & {\cellcolor{red!60} 20.45} & 0.8286                        & 17.29                        & 0.6457                        & 21.83                        & 0.9022                        \\
Histoformer \cite{sun2024restoring}                       & ECCV'24                                & Supervised                      & 3.51s                                     & 19.88                        & 0.8564 & {\cellcolor{red!60} 18.83} & 0.7849                        & {\cellcolor{orange!60} 22.36} & {\cellcolor{orange!60} 0.9259} \\
KA-Net \cite{feng2024advancing}                            & TPAMI'24                               & Supervised                      & 0.59s                                     & 17.59                        & 0.5440                        & 15.84                        & 0.5461                        & 17.93                        & 0.6691                        \\ \hline
DIP \cite{ulyanov2018deep}                               & CVPR'18                                & Unsupervised                    & 0.06s                                     & 13.75                        & 0.2506                        & 15.72                        & 0.4877                        & 21.21                        & 0.8163                        \\
DD \cite{heckel2018deep}                                & ICLR'19                                & Unsupervised                    & {\cellcolor{red!60} 0.02s}              & 17.40                        & 0.5095                        & 15.63                        & 0.5527                        & 21.91                        & 0.6450                        \\
DDIP \cite{gandelsman2019double}                              & CVPR'19                                & Unsupervised                    & 0.05s                                     & 12.92                        & 0.6856                        & 16.11                        & 0.6334                        & 16.24                        & 0.7714                        \\
ZID \cite{li2020zero}                               & TIP'20                                 & Unsupervised                    & 0.04s              & 16.21                        & 0.7487                        & 14.75                        & 0.7228                        & 16.29                        & 0.8211                        \\
YOLY \cite{li2021you}                              & IJCV'21                                & Unsupervised                    & 0.06s                                     & 16.53                        & 0.7752                        & 15.65                        & 0.7274                        & 17.48                        & 0.8575                        \\
USID-Net \cite{li2022usid}                          & TMM'21                                 & Unsupervised                    & 0.62s                                     & 12.68                        & 0.4665                        & 17.47                        & 0.5612                        & 19.10                        & 0.7047                        \\
D4 \cite{yang2022self}                                & CVPR'22                                & Unsupervised                    & 0.25s                                     & 18.64                        & 0.7892                        & 16.36                        & 0.6166                        & 16.60                        & 0.8517                        \\
ZS-NSN \cite{mansour2023zero}                      & CVPR'23    & Unsupervised      & {\cellcolor{orange!60} 0.03s}    & 19.98   & {\cellcolor{orange!60} 0.8855}  & 17.98 & 0.7900          & 21.95                    & 0.9211     \\
UCL-Dehaze \cite{wang2024ucl}                        & TIP'24                                 & Unsupervised                    & 0.07s                                     & 17.83                        & 0.6653                        & 15.04                        & 0.6221                        & 19.50                        & 0.7581                        \\
DIPDKP \cite{yang2024dynamic}                      & CVPR'24     & Unsupervised      & 0.54s   & 20.02   & 0.8227    & 18.36 & {\cellcolor{orange!60} 0.7927}    & 22.11    & 0.8725      \\
RSF-Dehaze                         & Ours                                    & Unsupervised                    & 0.16s                                     & {\cellcolor{orange!60} 20.13} & {\cellcolor{red!60} 0.8924} & {\cellcolor{orange!60} 18.59} & {\cellcolor{red!60} 0.8011} & {\cellcolor{red!60} 22.66} & {\cellcolor{red!60} 0.9445} \\ \hline

\label{tab1}
\end{tabular}}
\end{table*}

\begin{figure*}[h]
\centering
\includegraphics[width=\linewidth]{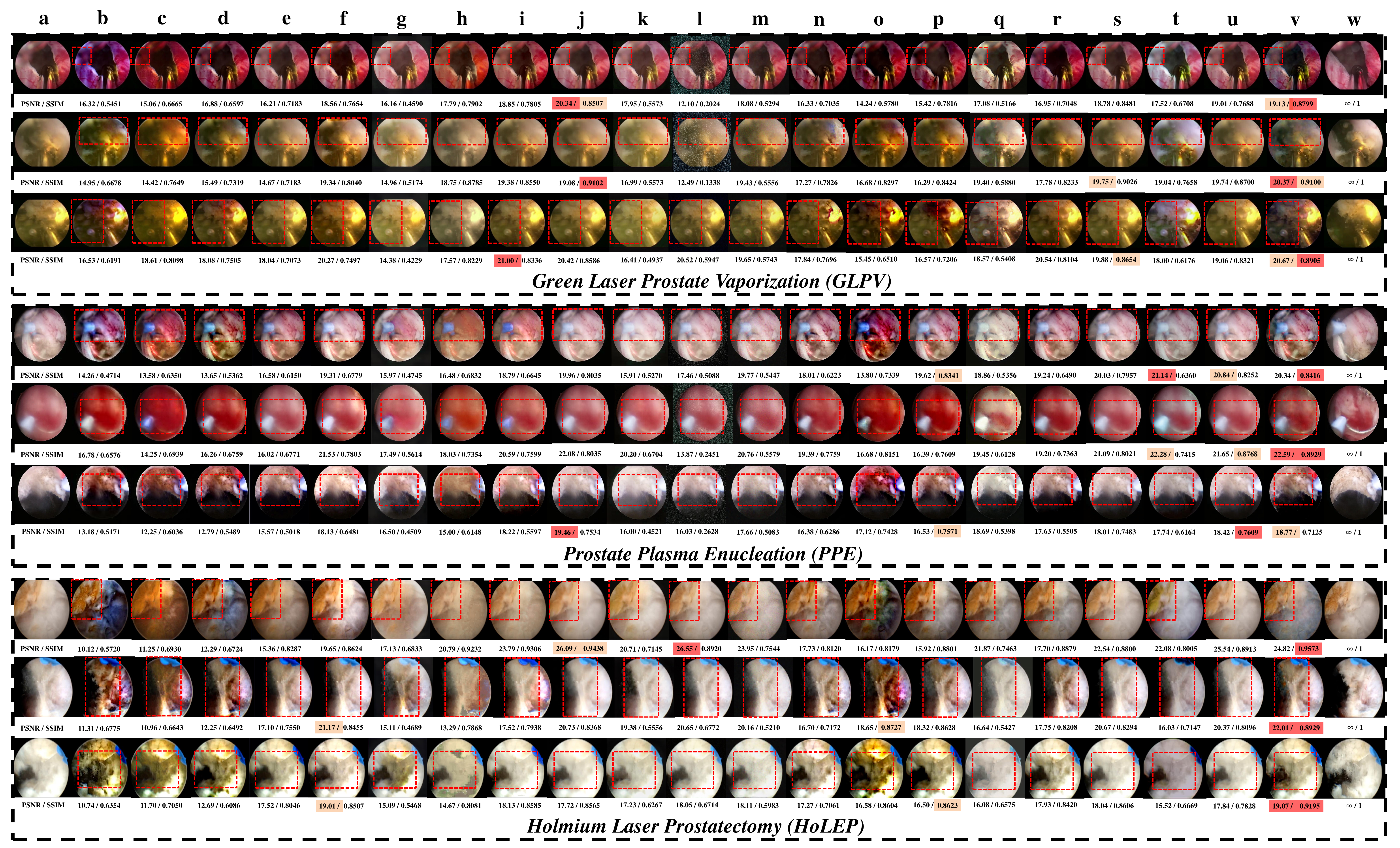}
\caption{Visualization of the dehazing results on USRobot-Dehaze dataset. The columns a-w represent the input blurred images, FVR, DCP, BCCR, AOD-Net, GCANet, 4KDehazing, Dehazeformer-t, Dehazeformer-m, Histoformer, KA-Net, DIP, DD, DDIP, ZID, YOLY, USID-Net, D4, ZS-NSN, UCL-Dehaze, DIPDKP, Ours and clean images, respectively.}
\label{fig04}
\end{figure*}

\section{EXPERIMENTS}
\subsection{Implementation Details}
All our experiments were implemented on the basis of Python 3.8 and Pytorch 1.13.0. Hardware support was provided by an NVIDIA GeForce RTX 4080 laptop GPU with a single memory of 12 GB. The number of iterations for a single image was set to 800. The optimizer used the ADAM optimizer and the learning rate was set to 0.001. The images were uniformly 1080$\times$1080. In addition, to keep the comparisons fair, we experimented with supervised methods for the baseline using the optimal model weights provided by the authors for either dehazing or image restoration. For evaluation metrics, we remained consistent with \cite{wang2024ucl,sun2024restoring,li2020zero,li2021you,feng2024advancing} and used the most widely used PSNR and SSIM. Higher values of these two metrics indicate better dehazing.

\subsection{State-of-the-Art Comparisons}
In order to validate the proposed RSF-Dehaze's dehazing performance for robotic vision in urological surgery, we conduct comparative experiments with 20 most advanced and popular dehazing and image recovery algorithms. As shown in Table \ref{tab1}, the quantitative evaluation results of the dehazing performed on the proposed USRobot-Dehaze dataset are presented. In particular, three different scenarios of the most common urological surgical robotic operations are included. From Table \ref{tab1}, it can be concluded that the proposed RSF-Dehaze outperforms both the most classical prior-based and the most advanced unsupervised methods. Although the proposed RSF-Dehaze has slightly lower PSNR metrics than the state-of-the-art supervised methods (Dehazeformer-M and Histoformer) for both GLPV and PPE scenes. However, our method is learned to dehaze only from a single image and does not have any other blurred image for pre-training. In particular, the proposed RSF-Dehaze inference time is reduced by 78.6$\%$ and 95.44$\%$ compared to Dehazeforme-M and Histoformer. In addition, we visualize the dehaze results under robotic vision for urological surgery in Fig. \ref{fig04}. By visualizing the dehazing result map, it can be concluded that the dehazing result map of the proposed RSF-Dehaze is able to be closer to the real value and the tissue structure is clearer and closer to the real situation. This is due to the fact that our proposed RSFM is able to repair the blurred organization region well. In particular, we further demonstrate this advantage in the next ablation experiments.

\begin{figure}[!t]
\centering
\includegraphics[width=\linewidth]{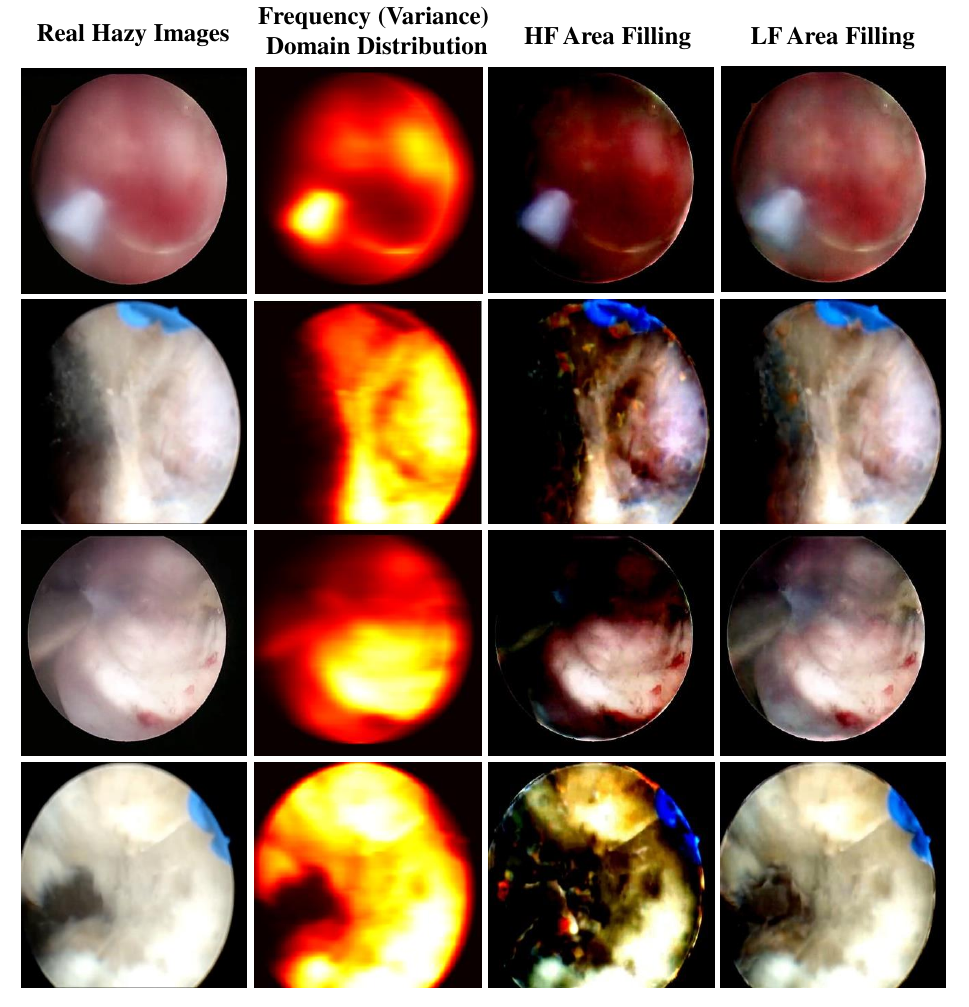}
\caption{Visualization of dehazing plots on RSFM with different filling methods.}
\label{fig05}
\end{figure}

\subsection{Ablation Study}
To further validate the effectiveness of the proposed regional similarity filling module (RSFM), we conduct ablation experiments. In particular, to highlight the effect of filling more, we fill the enhanced high-frequency information into the high-variance (frequency) region and the low-variance (frequency) region of the original image, respectively. Specifically, the experimental results and the visualized dehazing result plots are shown in Table \ref{tab2} and Fig. \ref{fig05}, respectively. From the visualization results, it can be concluded that the high variance (frequency) region becomes clearer while the low variance (frequency) region becomes blurrier when the enhanced high-frequency information by RSFM fills the high variance (frequency) region. From the quantitative results, both PSNR and SSIM showed a significant decrease in the three different urological surgical scenarios.

In addition, to verify the effectiveness of the reorganization operation and color conversion operation in the dehazing network of RSF-Dehaze, we performed ablation experiments, as shown in Table \ref{tab3}. We obtain a significant decrease in performance after the reorganization operation is removed. Specifically, PSNR metrics decreased by 9.6$\%$, 10.16$\%$, and 18.13$\%$ in the surgical scenarios of GLPV, PPE, and HoLEP, respectively, whereas SSIM metrics decreased by 8.46$\%$, 4.57$\%$, and 5.17$\%$, respectively. This reorganization operation enables the subsequent variance calculation of the RSFM to introduce less error. In addition, the dehazing performance of the model in all three surgical scenarios was degraded after we removed the color conversion operation. As can be seen in Table \ref{tab3}, both operations were effective in improving the model's ability to dehaze the three urological surgery scenarios. 

Moreover, to verify the performance of the dehazing network in RSF-Dehaze under robotic vision for urological surgery using different convolutional kernels, we perform ablation experiments in Table \ref{tab4}. We use convolution kernels of 3, 5, 7 and 9, respectively. In particular, we can conclude that changing the convolution kernel has little effect on the overall model performance. The best performance is achieved in both GLPV and HoLEP using a convolution kernel of 7, which is the default setting of the proposed convolution kernel for RSF-Dehaze. However, further use of convolution kernel of 9 shows a small decrease in the metrics.

\begin{table}[!t]
\centering
\caption{Ablation experiments on the effect of different filling methods of the proposed RSFM on the model performance.}
\renewcommand{\arraystretch}{1.5}
\resizebox{\linewidth}{!}{
\begin{tabular}{lcccccc}
\hline
                                    & \multicolumn{2}{c}{\textbf{GLPV}}                            & \multicolumn{2}{c}{\textbf{PPE}}                             & \multicolumn{2}{c}{\textbf{HoLEP}}                           \\ \cline{2-7} 
\multirow{-2}{*}{\textbf{Settings}} & \textbf{PSNR}                & \textbf{SSIM}                 & \textbf{PSNR}                & \textbf{SSIM}                 & \textbf{PSNR}                & \textbf{SSIM}                 \\ \hline
HF Area Filling                     & 18.34                        & 0.8276                        & 17.74                        & 0.7701                        & 20.84                        & 0.9251                        \\
LF Area Filling                     & {\cellcolor{red!60} 20.13} & {\cellcolor{red!60} 0.8924} & {\cellcolor{red!60} 18.59} & {\cellcolor{red!60} 0.8011} & {\cellcolor{red!60} 22.66} & {\cellcolor{red!60} 0.9445} \\ \hline

\label{tab2}
\end{tabular}}
\end{table}

\begin{table}[!t]
\centering
\caption{Ablation experiments on the effect of reorganization and color conversion operations on the model performance.}
\renewcommand{\arraystretch}{1.5}
\resizebox{\linewidth}{!}{
\begin{tabular}{lcccccc}
\hline
                                    & \multicolumn{2}{c}{\textbf{GLPV}}                            & \multicolumn{2}{c}{\textbf{PPE}}                             & \multicolumn{2}{c}{\textbf{HoLEP}}                           \\ \cline{2-7} 
\multirow{-2}{*}{\textbf{Settings}} & \textbf{PSNR}                & \textbf{SSIM}                 & \textbf{PSNR}                & \textbf{SSIM}                 & \textbf{PSNR}                & \textbf{SSIM}                 \\ \hline
RSF-Dehaze (Ours)                   & {\cellcolor{red!60} 20.13} & {\cellcolor{red!60} 0.8924} & {\cellcolor{red!60} 18.59} & {\cellcolor{red!60} 0.8011} & {\cellcolor{red!60} 22.66} & {\cellcolor{red!60} 0.9445} \\ \hline
w/o Reorganization                  & 18.19                        & 0.8169                        & 16.70                         & {\cellcolor{orange!60} 0.7645}                        & 18.55                        & 0.8957                        \\
w/o Color Conversion & {\cellcolor{orange!60} 20.01} & {\cellcolor{orange!60} 0.8786}  & {\cellcolor{orange!60} 18.20} & 0.7545 & {\cellcolor{orange!60} 22.44} & {\cellcolor{orange!60} 0.9405} \\ \hline
\label{tab3}
\end{tabular}}
\end{table}

\begin{table}[!t]
\centering
\caption{Ablation experiments using the effect of convolution with different convolution kernels on model performance.}
\renewcommand{\arraystretch}{1.5}
\resizebox{\linewidth}{!}{
\begin{tabular}{lcccccc}
\hline
                                              & \multicolumn{2}{c}{\textbf{GLPV}}                            & \multicolumn{2}{c}{\textbf{PPE}}                             & \multicolumn{2}{c}{\textbf{HoLEP}}                           \\ \cline{2-7} 
\multirow{-2}{*}{\textbf{Convolution Kernel}} & \textbf{PSNR}                & \textbf{SSIM}                 & \textbf{PSNR}                & \textbf{SSIM}                          & \textbf{PSNR}                & \textbf{SSIM}                 \\ \hline
3$\times$3                                           & 20.09                        & {\cellcolor{orange!60} 0.8920}  & {\cellcolor{red!60} 18.67} & {\cellcolor{red!60} 0.8014} & 22.6                         & 0.9441                        \\
5$\times$5                                           & {\cellcolor{orange!60} 20.11} & 0.8897                        & 18.62                        & 0.8004                        & 22.55                        & {\cellcolor{orange!60} 0.9442} \\
7$\times$7                                           & {\cellcolor{red!60} 20.13} & {\cellcolor{red!60} 0.8924} & 18.59                        & {\cellcolor{orange!60} 0.8011} & {\cellcolor{red!60} 22.66} & {\cellcolor{red!60} 0.9445} \\
9$\times$9                                           & 20.08                        & 0.8889                        & {\cellcolor{orange!60} 18.60} & 0.8008                        & {\cellcolor{orange!60} 22.65} & 0.9435                        \\ \hline

\label{tab4}
\end{tabular}}
\end{table}

\section{CONCLUSION}
In this paper, we propose an unsupervised zero-shot dehaze method (RSF-Dehaze) for robotic vision in urological surgery. For the fogging characteristics of liquids in urological surgery robotic vision, a region similarity filling module (RSFM) is proposed to significantly improve the recovery of blurred regional tissues. In particular, the proposed RSF-Dehaze is an unsupervised zero-shot dehazing method, which does not rely on any additional training data to obtain better dehazing results. In addition, we also organize and propose a dehazing (USRobot-Dehaze) dataset for robotic vision in urological surgery. To the best of our knowledge, we are the first team to propose a publicly available dehaze dataset for robotic vision in urological surgery. The USRobot-Dehaze dataset contains the three most common scenarios of urological surgical robot operation. Extensive experiments conducted across the three scenarios of our proposed USRobot-Dehaze dataset demonstrate that RSF-Dehaze achieves superior results, both quantitatively and qualitatively.


\end{document}